\documentclass[runningheads]{llncs}
\usepackage[T1]{fontenc}
\usepackage{graphicx,verbatim}
\usepackage{graphicx}
\usepackage{color}
\usepackage{subcaption}
\usepackage{amsmath}
\usepackage{amssymb}
\usepackage{hyperref}
\usepackage{bbding}
\usepackage{multirow} 

\usepackage[bottom]{footmisc}

\usepackage{color}

\newcommand{\papername}{NerT-CA}

\begin{document}
\title{\papername: Efficient Dynamic Reconstruction from Sparse-view X-ray Coronary Angiography}

\titlerunning{\papername}

\author{Kirsten W.H. Maas\inst{1}\orcidID{0009-0007-4402-0301} \and
Danny Ruijters\inst{2, 3}\orcidID{0000-0002-9931-4047} \and
Nicola Pezzotti\inst{1, 2}\orcidID{0000-0001-9554-4331} \and 
Anna Vilanova\inst{1}\orcidID{0000-0002-1034-737X}}
\authorrunning{K.W.H. Maas et al.}
%
\institute{Department of Mathematics and Computer Science, Eindhoven University of Technology, Eindhoven, The Netherlands \\
\email{\{k.w.h.maas, a.vilanova\}@tue.nl}\\
\email{nicola.pezzotti@proton.me}
\and
Philips Healthcare, Best, The Netherlands\\
\email{danny.ruijters@philips.com}
\and
Department of Electrical Engineering, Eindhoven University of Technology, The Netherlands
}


\maketitle              
\begin{abstract} 
Three-dimensional (3D) and dynamic 3D+time (4D) reconstruction of coronary arteries from X-ray coronary angiography (CA) has the potential to improve clinical procedures.
However, there are multiple challenges to be addressed, most notably, blood-vessel structure sparsity, poor background and blood vessel distinction, sparse-views, and intra-scan motion.
State-of-the-art reconstruction approaches rely on time-consuming manual or error-prone automatic segmentations, limiting clinical usability. 
Recently, approaches based on Neural Radiance Fields (NeRF) have shown promise for automatic reconstructions in the sparse-view setting.
However, they suffer from long training times due to their dependence on MLP-based representations.
We propose \papername, a hybrid approach of Neural and Tensorial representations for accelerated 4D reconstructions with sparse-view CA.
Building on top of the previous NeRF-based work, we model the CA scene as a decomposition of low-rank and sparse components, utilizing fast tensorial fields for low-rank static reconstruction and neural fields for dynamic sparse reconstruction.
Our approach outperforms previous works in both training time and reconstruction accuracy, yielding reasonable reconstructions from as few as three angiogram views.
We validate our approach quantitatively and qualitatively on representative 4D phantom datasets.


\end{abstract}

\section{Introduction}

X-ray coronary angiography (CA) remains the gold standard for diagnosing and treating coronary artery disease~\cite{ccimen2016reconstruction}.
It captures the coronary arteries as a dynamic sequence of two-dimensional (2D) X-ray projections, referred to as coronary angiograms.
These angiograms inherently represent four-dimensional (4D) information, as they capture the dynamic 3D motion of the coronary arteries largely induced by cardiac and respiratory motion~\cite{ccimen2016reconstruction}.
However, since these X-rays provide only 2D projections of 3D structure, depth foreshortening occurs, which can negatively impact the perception of the vessel structure.
To mitigate this, multiple coronary angiograms are acquired from different viewpoints during interventions~\cite{green2004three}. 
Yet, to avoid X-ray exposure, typically no more than four views are captured~\cite{di2005coronary}.
A 4D reconstruction could improve CA procedures by providing a comprehensive view of the coronary arteries, aiding, for example, in catheter navigation for roadmap overlays while limiting the X-ray exposure~\cite{piayda2018dynamic}.

Even though clinically relevant, the problem of 4D reconstruction from CA data remains challenging.
CA data exhibits several complex properties, namely the extremely sparse-view setting, intra-scan motion, and the intricate characteristics of the blood vessel, such as structure sparsity, overlapping background structures, and contrast inhomogeneity~\cite{ccimen2016reconstruction}.
Traditional successful reconstruction methods rely on sequences acquired from many viewpoints or time-consuming manual segmentations, not fitting clinical workflows~\cite{ccimen2016reconstruction,iyer2023multi,green2004three}.
Recent approaches utilize automatic segmentation models to reconstruct from two angiogram sequences, but suffer from noise propagated from the segmentation errors~\cite{hwang2021simple,bappy2021automated,zhu2025sparse}.

\begin{figure}[!b]
\centering
\includegraphics[width=0.9\textwidth]{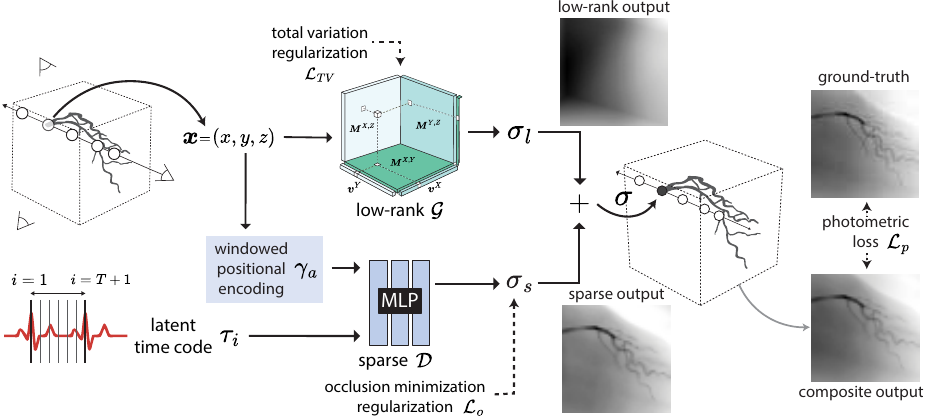}
\caption{\papername \space models the dynamic CA scene as a self-supervised decomposition of low-rank and sparse components. The low-rank component is modeled by a tensorial field and the sparse component by a dynamic neural field.}
\label{fig:overview}
\end{figure}

Approaches building on Neural Radiance Fields (NeRFs)~\cite{mildenhall2021nerf} have been proposed, which directly reconstruct from the x-ray images, avoiding reliance on segmentation models~\cite{maas2024nerf,kshirsagar2024generative}.
NeRFs are powerful for 3D and 4D reconstruction by learning individual continuous scenes from a limited number of projections~\cite{mildenhall2021nerf,yang2023freenerf}.
Most notably, NeRF-CA~\cite{maas2024nerf} recently demonstrated reasonable reconstructions from four synthetic coronary angiograms exhibiting cardiac motion.
They decompose the CA scene into a static and dynamic component, each modeled by a multi-layer perceptron (MLP).
While significantly outperforming state-of-the-art X-ray NeRF-based techniques in reconstruction quality, their method suffers from hour-long training times.
Meanwhile, many acceleration methods for general NeRFs have been proposed~\cite{zha2025r,chen2022tensorf}, which could potentially be leveraged to accelerate NeRF-CA.
For example, Tensorial Radiance Field (TensoRF)~\cite{chen2022tensorf} models scenes through fast tensorial decompositions~\cite{kolda2009tensor}, which has been shown beneficial to replace MLP-based representations in medical settings~\cite{yang2024efficient}, but are not directly applicable to the CA setting.

Inspired by these developments, we present \papername, a hybrid approach of Neural and Tensorial representations that efficiently reconstructs dynamic CA scenes from as few as three coronary angiogram views.
Building on the foundation of NeRF-CA, we decompose the CA scene into static and dynamic components.
In contrast, we extend this decomposition by applying low-rank and sparse decomposition constraints~\cite{gao2011robust}, utilizing a combination of neural and tensorial representations with tailored regularizations. 
Specifically, the low-rank component is modeled as an efficient tensorial field with smoothness regularization.
Concurrently, the sparse component is represented as a dynamic neural field, where we impose specific regularization for the sparse-view setting.
Not only does this approach gain a significant speed-up, but we also show that it benefits reconstruction quality in the sparse-view setting, outperforming NeRF-CA and state-of-the-art X-ray reconstruction techniques.
We evaluate our work quantitatively and qualitatively using representative 4D phantoms~\cite{segars20104d,rosset2004osirix}.

\section{Method}

\papername \space is built on top of NeRF-CA~\cite{maas2024nerf}, a method that utilizes static and dynamic decomposition to reconstruct synthetic CA scenes with four coronary angiograms that exhibit cardiac motion.
We propose a hybrid approach, combining fast tensorial fields~\cite{chen2022tensorf} with dynamic neural fields~\cite{mildenhall2021nerf,park2021nerfies}, to accelerate this method while improving reconstruction quality. 
Figure~\ref{fig:overview} provides an overview of \papername.
We will detail each element in the next sections.
The total loss function is given by
\begin{equation*}
    \mathcal{L} = \mathcal{L}_p + \lambda_{TV}\mathcal{L}_{TV} + \lambda_{o}\mathcal{L}_o,
\end{equation*}
where $\mathcal{L}_p$, $\mathcal{L}_{TV}$, and $\mathcal{L}_o$ are the photometric, total variation (TV), and occlusion minimization losses, respectively, with their according weights $\lambda_{TV}$ and $\lambda_o$.
Compared to NeRF-CA, this loss term is simplified by removing several regularization techniques, but introducing the $\mathcal{L}_{TV}$ term enabled by the tensorial field.

\subsection{Neural and Tensorial Representations}
Neural Radiance Fields (NeRFs) encode a continuous scene through a multilayer perceptron (MLP)~\cite{mildenhall2021nerf}. 
For the X-ray setting~\cite{ruckert2022neat}, NeRF-CA learns a function $\mathcal{F}_{\Theta}(\boldsymbol{x}) = \sigma$, where $\boldsymbol{x} \in \mathbb{R}^3$ is a 3D spatial coordinate, $\sigma$ is the attenuation coefficient, and $\Theta$ represents the weights of the MLP.
To model high-frequency details, NeRF-CA applies positional encoding $\gamma$ to map inputs to a higher-dimensional space, a common approach in NeRFs~\cite{tancik2020fourier,mildenhall2021nerf}. 
A dynamic scene is modeled as $\mathcal{D}_{\Theta}(\boldsymbol{x}, \tau_i) = \sigma$, where a learnable time latent-code $\tau_i$ represents each unique time $i$ in the scene~\cite{park2021nerfies}.
In NeRF-CA and our method, $\tau_i$ represents discrete cardiac phases $i \in {1, \dots, T}$ extracted from an ECG signal, as illustrated in Figure~\ref{fig:overview}.
While powerful for high-fidelity scene reconstruction, NeRF-CA, like other NeRF architectures~\cite{muller2022instant}, suffers from high computational costs due to the dense sampling of 3D points needed to be evaluated by the MLP.

Tensorial Radiance Field (TensoRF) addresses the computational costs by introducing an explicit volumetric representation of the attenuation field~\cite{chen2022tensorf}.
Instead of utilizing MLPs, TensoRF decomposes the volumetric space using low-rank tensor decomposition~\cite{kolda2009tensor}. 
The 3D volume is represented as a tensor $\mathcal{G} \in \mathbb{R}^{D \times D \times D}$, where $D$ represents the resolution of the grid in the dimensions $X, Y, $ and $Z$.
TensoRF specifically proposes vector-matrix (VM) decomposition, where $\mathcal{G}$ is decomposed with a component that consists of orthogonal pairs of learnable vectors $\boldsymbol{v}_r^X, \boldsymbol{v}_r^Y, \boldsymbol{v}_r^Z \in \mathbb{R}^D$ and learnable matrices $\boldsymbol{M}^{Y, Z}_r, \boldsymbol{M}^{X, Z}_r, \boldsymbol{M}^{X, Y}_r \in \mathbb{R}^{D \times D}$.
A number of these components or ranks $r \in \{0, \dots, R\}$, where $R$ represents the maximum rank, are then used to approximate the tensor $\mathcal{G}$, where a higher $R$ allows for more detailed scenes to be modeled.
Specifically, $\mathcal{G}$ is represented as
\begin{equation*}
    \mathcal{G} = \sum_{r=1}^R \boldsymbol{v}_r^X \circ \boldsymbol{M}_r^{Y, Z} + \boldsymbol{v}^Y_r \circ \boldsymbol{M}_r^{X, Z} + \boldsymbol{v}^Z_r \circ \boldsymbol{M}_r^{X, Y},
\end{equation*}
where $\circ$ represents the outer product. 
Evaluating $\mathcal{G}(\boldsymbol{x}) = \sigma$ is performed by retrieving its corresponding values from the vectors and matrices, as schematically shown in Figure~\ref{fig:overview}.
The values of these vectors and matrices are optimized with gradient descent.
Trilinear interpolation within these vectors and matrices is applied to achieve continuous outputs.
The advantage of this tensorial representation, in comparison to the neural representation, is that fewer parameters need to be learned for the same value~\cite{chen2022tensorf}.

\subsection{Low-rank and Sparse Decomposition}
In this work, we model the dynamic CA scene as a mixture of low-rank and sparse (L+S) components~\cite{gao2011robust}.
Low-rank components are assumed to be stationary and low-frequency, whereas sparse components are assumed to be dynamic with minimal non-zero attenuation values.
Given a CA scene as $\mathcal{H}(\boldsymbol{x}, \tau_i)$, we model it as a decomposition of low-rank stationary field $\mathcal{G}(\boldsymbol{x}) = \sigma_l$ and sparse dynamic field $\mathcal{D}_{\Theta}(\boldsymbol{x}, \tau_i) = \sigma_s$ as
\begin{equation*}
    \mathcal{H}(\boldsymbol{x}, \tau_i) = \mathcal{G}(\boldsymbol{x}) + \mathcal{D}_{\Theta}(\gamma_a(\boldsymbol{x}), \tau_i) = \sigma_l + \sigma_s.
\end{equation*}

Prior work on 4D medical scenes has shown that L+S decomposition enforces temporal coherence by sparsifying dynamic components, enabling accurate separation of background and motion~\cite{gao2011robust}.
NeRF-CA also decomposed scenes into static and dynamic parts, but via a hard factorization that forces the points to be static or dynamic~\cite{maas2024nerf}.
Given the sparsity of coronary structures, this can lead to missing vessels in reconstructions.
We propose an L+S decomposition as it better reflects CA scenes, where contrast injection naturally produces a composition of background and blood vessel structure.

To enforce the L+S properties, we model the low-rank component as a static tensorial field $\mathcal{G}(\boldsymbol{x})$ and the sparse component as a dynamic neural field $\mathcal{D}_{\Theta}(\gamma_a(\boldsymbol{x}), \tau_i)$.
We use a tensorial field as it is computationally fast and can be regularized to be smooth globally by using a small maximum rank $R$.
Meanwhile, the sparse component is modeled through a dynamic neural $\mathcal{D}_{\Theta}(\gamma_a(\boldsymbol{x}), \tau_i)$ to model the coronary artery, similarly to NeRF-CA. 

Given $\mathcal{H}(\boldsymbol{x}, \tau_i) = \sigma_l + \sigma_s$, we compute the predicted pixel intensities per cardiac phase $i$ using the discretized Beer-Lambert law~\cite{max2002optical}.
The photometric loss $\mathcal{L}_p$ is then defined as the mean-squared error (MSE) between predicted and ground-truth frame intensities, following NeRF-CA.

\subsection{Low-rank and Sparse-view Regularizations}

To enable L+S decomposition in the sparse-view setting, we apply regularizations to both the tensorial and neural fields, as shown in Figure~\ref{fig:overview}.
For the smooth low-rank tensorial field, we utilize total variation (TV) regularization $\mathcal{L}_{TV}$ on the learned matrices and vectors directly, similarly to TensoRF~\cite{chen2022tensorf}.

For the neural field, we apply the same two regularization techniques as in NeRF-CA, windowed positional encoding and occlusion minimization, to prevent degenerate solutions in the sparse-view setting.
Windowed positional encoding $\gamma_a$ introduces a coarse-to-fine training schedule to suppress high-frequency artifacts, which is implemented through an annealing schedule on the frequency encoding~\cite{yang2023freenerf,wu2022d}.
Occlusion minimization $\mathcal{L}_o$ penalizes density near the X-ray source to prevent floaters near the camera as a masked L1 regularization on the attenuation coefficients~\cite{yang2023freenerf}.
We refer to NeRF-CA~\cite{maas2024nerf} for further details.

\section{Experimental Setup}

\noindent \textbf{Datasets} \hspace{4pt} We evaluate on 4D phantom datasets exhibiting cardiac motion, allowing for quantitative evaluation~\cite{maas2024nerf,rohkohl2010cavarev}. 
We use the XCAT~\cite{segars20104d} and MAGIX~\cite{rosset2004osirix} phantoms to generate realistic CA data of the left coronary artery (LCA).
XCAT is a parameterized 4D phantom sampled at 0.5 mm$^3$ isotropic resolution~\cite{segars20104d} as $T = 10$ volumes.
MAGIX is a 4D CCTA dataset at $0.4\times0.4\times2$ mm$^3$ resolution with $T = 10$ volumes.
Like NeRF-CA~\cite{maas2024nerf}, we utilize the tomographic toolbox TIGRE~\cite{biguri2016tigre} to generate realistic CA projections of $200 \times 200$ pixels for the single-plane C-arm setting with standard clinical angles~\cite{di2005coronary}.
NeRF-CA defines this training setting for a minimum of four projections with four validation projections.
We use the same setting, where for our three-view training setting, we exclude the projection with the most blood vessel overlap.\\\\
\noindent \textbf{Evaluation metrics} \hspace{4pt} We perform quantitative and qualitative evaluations. 
The qualitative approach is necessary to overcome challenges in evaluating medical imaging with standard quantitative metrics~\cite{kastryulin2023image}.
Quantitatively, we focus on 2D novel view synthesis for clinical roadmap overlays~\cite{piayda2018dynamic}.
Our evaluation focuses on the masked blood vessel Dice score (DSC), evaluating the overlap between the segmented blood vessel structure and the ground-truth, following NeRF-CA.
This DSC score is computed by thresholding the maximum intensity projection of the predicted attenuation values.
Since our focus is blood vessel reconstruction, DSC is a more suitable metric, as also demonstrated in NeRF-CA. 
For completeness, we additionally report common image quality metrics, namely peak signal-to-noise ratio (PSNR) and structural similarity (SSIM).
We also discuss training times in minutes and inference times in frames per second (FPS).
The image quality metrics are reported as the mean of 40 views, which are the 4 validation views across all time steps $T = 10$.\\\\
\noindent \textbf{Implementation details} \hspace{4pt} The models were implemented in PyTorch and trained on an RTX A5000 GPU for $30,000$ iterations using a ray batch size of $2048$ and $256$ samples per ray.
Hyperparameters were selected via grid search. 
The tensorial field uses a grid size of $D = 48$ and maximum rank $R = 3$, while the neural field has $4$ layers of $128$ ReLU neurons. 
A softplus activation ensures positive attenuation outputs.
We use the Adam optimizer with a linear learning rate schedule: $10^{-2}$ to $10^{-3}$ for the tensorial field and $10^{-1}$ to $10^{-2}$ for the neural field. 
To let the tensorial field first capture coarse background, neural field optimization is delayed by $1500$ iterations.
Hyperparameters are consistent across datasets, except for the total variation weight $\lambda_{TV}$, with $\lambda_{TV} = 10^{-3}$ for the MAGIX dataset and $\lambda_{TV} = 10^{-2}$ for the XCAT dataset, adjusted for the finer resolution of the XCAT dataset.
We impose a maximum frequency band for the windowed positional encoding of $10$ over $15,000$ iterations. 
The occlusion minimization weight is $\lambda_o = 10^{-8}$, with the same distance threshold as NeRF-CA.
The code repository of NerT-CA is publicly available~\footnote[1]{\url{https://github.com/kirstenmaas/NerT-CA}}.\\\\
\noindent \textbf{Baselines} \hspace{4pt} State-of-the-art CA reconstruction methods require input segmentations for reconstructions~\cite{fu20243dgr,zhu2025sparse}.
As our method is fully automatic, we compare it to state-of-the-art sparse-view reconstruction methods for X-ray data, besides the dynamic NeRF-CA.
These baselines include SAX-NeRF~\cite{cai2024structure}, X-Gaussian~\cite{cai2024radiative}, and R2-Gaussian~\cite{zha2025r} with default parameters, which all model static X-ray data.
The latter two were introduced after NeRF-CA, demonstrating reconstructions from a minimum of $25$ views.
For these methods, DSC is computed from the directly predicted attenuation values, whereas for our method and NeRF-CA, they are computed from the dynamic attenuation values.

\section{Results and Discussion}

\begin{table}[!t]
\caption{Quantitative results for the XCAT and MAGIX datasets for 3, 4, and 9 training projections. Bolding is used to indicate the best score.}\label{tab1}
\renewcommand{\arraystretch}{1.2}
\fontsize{8pt}{8pt}\selectfont
\centering
\begin{tabular}{l|l|ccc|ccc|ccc}
\hline
\multicolumn{1}{c|}{\multirow{2}{*}{Dataset}} & \multicolumn{1}{c|}{\multirow{2}{*}{Method}} & \multicolumn{3}{c|}{3-view}                                                              & \multicolumn{3}{c|}{4-view}                                                              & \multicolumn{3}{c}{9-view}                                                               \\ \cline{3-11} 
\multicolumn{1}{c|}{}                         & \multicolumn{1}{c|}{}                        & \multicolumn{1}{c|}{DSC}           & \multicolumn{1}{c|}{PSNR}           & SSIM          & \multicolumn{1}{c|}{DSC}           & \multicolumn{1}{c|}{PSNR}           & SSIM          & \multicolumn{1}{c|}{DSC}           & \multicolumn{1}{c|}{PSNR}           & SSIM          \\ \hline
\multicolumn{1}{c|}{}                         & SAX-NeRF                                     & \multicolumn{1}{c|}{0.00}          & \multicolumn{1}{c|}{13.49}          & 0.58          & \multicolumn{1}{c|}{0.00}          & \multicolumn{1}{c|}{12.89}          & 0.58          & \multicolumn{1}{c|}{0.00}          & \multicolumn{1}{c|}{11.44}          & 0.55          \\
                                              & X-Gaussian                                   & \multicolumn{1}{c|}{0.05}          & \multicolumn{1}{c|}{12.58}          & 0.49          & \multicolumn{1}{c|}{0.22}          & \multicolumn{1}{c|}{12.53}          & 0.51          & \multicolumn{1}{c|}{0.37}          & \multicolumn{1}{c|}{9.30}           & 0.49          \\
\multicolumn{1}{c|}{XCAT}                     & R2-Gaussian                                  & \multicolumn{1}{c|}{0.01}          & \multicolumn{1}{c|}{13.02}          & 0.57          & \multicolumn{1}{c|}{0.00}          & \multicolumn{1}{c|}{12.87}          & 0.58          & \multicolumn{1}{c|}{0.49}          & \multicolumn{1}{c|}{15.77}          & 0.71          \\
                                              & NeRF-CA                                      & \multicolumn{1}{c|}{0.41}          & \multicolumn{1}{c|}{11.69}          & 0.65          & \multicolumn{1}{c|}{0.76}          & \multicolumn{1}{c|}{11.66}          & 0.72          & \multicolumn{1}{c|}{0.74}          & \multicolumn{1}{c|}{\textbf{16.25}} & 0.83          \\
                                              & Ours                                         & \multicolumn{1}{c|}{\textbf{0.75}} & \multicolumn{1}{c|}{\textbf{15.34}} & \textbf{0.79} & \multicolumn{1}{c|}{\textbf{0.84}} & \multicolumn{1}{c|}{\textbf{15.16}} & \textbf{0.80} & \multicolumn{1}{c|}{\textbf{0.87}} & \multicolumn{1}{c|}{14.12}          & \textbf{0.84} \\ \hline
\multicolumn{1}{c|}{}                         & SAX-NeRF                                     & \multicolumn{1}{c|}{0.00}          & \multicolumn{1}{c|}{13.70}          & 0.71          & \multicolumn{1}{c|}{0.00}          & \multicolumn{1}{c|}{14.17}          & 0.72          & \multicolumn{1}{c|}{0.00}          & \multicolumn{1}{c|}{13.72}          & 0.74          \\
                                              & X-Gaussian                                   & \multicolumn{1}{c|}{0.26}          & \multicolumn{1}{c|}{13.73}          & 0.65          & \multicolumn{1}{c|}{0.20}          & \multicolumn{1}{c|}{14.54}          & 0.68          & \multicolumn{1}{c|}{0.47}          & \multicolumn{1}{c|}{14.20}          & 0.67          \\
\multicolumn{1}{c|}{MAGIX}                    & R2-Gaussian                                  & \multicolumn{1}{c|}{0.00}          & \multicolumn{1}{c|}{13.09}          & 0.55          & \multicolumn{1}{c|}{0.00}          & \multicolumn{1}{c|}{13.90}          & 0.58          & \multicolumn{1}{c|}{0.00}          & \multicolumn{1}{c|}{16.24}          & 0.70          \\
                                              & NeRF-CA                                      & \multicolumn{1}{c|}{0.82}          & \multicolumn{1}{c|}{\textbf{18.46}} & 0.80          & \multicolumn{1}{c|}{0.81}          & \multicolumn{1}{c|}{13.34}          & 0.72          & \multicolumn{1}{c|}{0.90}          & \multicolumn{1}{c|}{18.49}          & 0.85          \\
                                              & Ours                                         & \multicolumn{1}{c|}{\textbf{0.88}} & \multicolumn{1}{c|}{14.97}          & \textbf{0.82} & \multicolumn{1}{c|}{\textbf{0.90}} & \multicolumn{1}{c|}{\textbf{16.70}} & \textbf{0.86} & \multicolumn{1}{c|}{\textbf{0.90}} & \multicolumn{1}{c|}{\textbf{18.92}} & \textbf{0.88} \\ \hline
\end{tabular}%
\end{table}

\begin{figure}[!t]
\centering
\includegraphics[width=\textwidth]{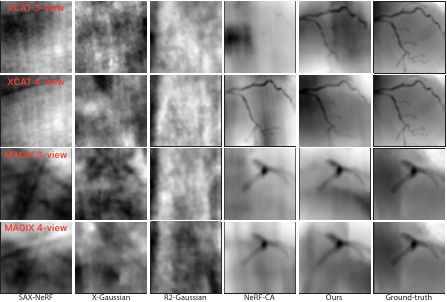}
\caption{Qualitative results for XCAT (3 and 4-view) and MAGIX (3 and 4-view).}
\label{fig:qualitative}
\end{figure}

We report reconstruction quality across different training view settings: 3, 4, and 9 projections.
Table~\ref{tab1} shows DSC, PSNR, and SSIM scores for XCAT and MAGIX.
Our method consistently outperforms all baselines in DSC, particularly in the 3-view setting.
Except for NeRF-CA, baselines fail to reconstruct vessel structures, reflected in low DSC scores.
Our method generally achieves better PSNR and SSIM, with occasional exceptions where NeRF-CA scores higher PSNR, likely due to background oversmoothing from TV regularization.
Figure~\ref{fig:qualitative} shows qualitative results at time point $i = 1$.
The baselines yield degenerate reconstructions due to background distractions, except NeRF-CA.
Compared to NeRF-CA, our method recovers finer vessels for 4 projections and robustly reconstructs in the 3-view setting on XCAT, with comparable quality on MAGIX.

\begin{figure}[!t]
\centering
\includegraphics[width=\textwidth]{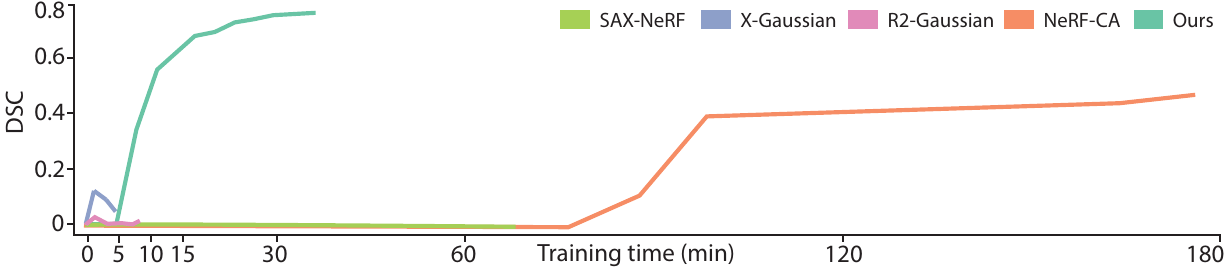}
\caption{Dice (DSC) scores over training time in minutes for the XCAT 3 training view setting for our method and the baselines.}
\label{fig:time}
\end{figure}

We also analyze the evolution of DSC score during training for the 3-view XCAT setting compared to the baselines, as shown in Figure~\ref{fig:time}.
SAX-NeRF~\cite{cai2024structure} fails to capture a blood vessel structure within one hour, leading to a zero DSC score.
The Gaussian-based baselines~\cite{cai2024radiative,zha2025r} converge within minutes, but they fail to produce reconstructions, as reflected by their low DSC scores.
NeRF-CA~\cite{maas2024nerf} achieves a higher DSC score but requires hours-long training time.
Notably, our method outperforms all baselines in DSC score with a running time of 37 minutes.
For the other dataset and projection settings, we observe similar trends.

Finally, we report detailed training and inference times. 
Since only NeRF-CA and our method capture vessel structures, we report only these two. 
On average, NeRF-CA requires 361 minutes for training and runs at $0.5$ FPS, while our method trains in 37 minutes and runs at $2$ FPS.
Therefore, we achieve an approximate $10\times$ speedup in training and $4\times$ in inference, while also improving reconstruction quality.\\

\noindent \textbf{Ablation study} \hspace{4pt} We ablate the effect of the regularization techniques, total variation $\mathcal{L}_{TV}$, windowed positional encoding $\gamma_a$, and occlusion minimization $\mathcal{L}_o$, in Table~\ref{tab2}.
Enforcing all regularizations yields the highest DSC scores, highlighting their importance and robustness across both datasets.
For XCAT, the regularizations significantly improve vessel reconstruction, while for the lower-resolution MAGIX dataset, improvements are smaller.

\begin{table}[!t]
\centering
\caption{Ablation of the regularizations in three-view training setting: total variation $\mathcal{L}_{TV}$, windowed positional encoding $\gamma_a$, and occlusion minimization $\mathcal{L}_o$.}\label{tab2}
\fontsize{8pt}{8pt}\selectfont
\renewcommand{\arraystretch}{1.1}
\begin{tabular}{ccc|ccc|ccc}
\hline
                            &                           &                             & \multicolumn{3}{c|}{XCAT}                                                                                              & \multicolumn{3}{c}{MAGIX}                                                                                              \\ \hline
$\mathcal{L}_{TV}$                 & $\gamma_a$               & $\mathcal{L}_o$                 & DSC                         & PSNR                         & SSIM                        & DSC                         & PSNR                         & SSIM                        \\ \hline
\XSolidBrush & \Checkmark & \XSolidBrush & 0.09          & 13.97          & 0.69          & 0.84          & \textbf{15.65} & 0.70          \\
\Checkmark   & \Checkmark & \XSolidBrush & 0.58    & 11.82          & 0.74    & 0.54          & 15.24    & 0.80    \\
\XSolidBrush & \Checkmark & \Checkmark   & 0.58   & 14.32    & 0.71          & 0.86    &      14.37          & 0.67          \\ \hline
\Checkmark   & \Checkmark & \Checkmark   & \textbf{0.75} & \textbf{15.34} & \textbf{0.79} & \textbf{0.88} & 14.97          & \textbf{0.82} \\ \hline
\end{tabular}%
\end{table}

\section{Conclusion}

We propose \papername, an efficient 4D reconstruction method for synthetic CA scenes from as few as three angiogram views. 
Our method extends a prior NeRF-based approach by modeling the scene as a low-rank and sparse decomposition using a combination of tensorial and neural fields with dedicated regularizations. 
Experiments demonstrate that \papername \space reduces training time by a factor of $10$ while significantly improving reconstruction quality in sparse-view settings.
We conduct an ablation study to validate the importance of our regularization components.
Future work could explore integrating 3D Gaussian Splatting~\cite{kerbl20233d} into our decomposition framework for further acceleration.

\begin{credits}
\subsubsection{\ackname} This research was performed within the Medusa project as part of the Eindhoven MedTech Innovation Center research collaboration between Eindhoven University of Technology, Philips Healthcare, and Catharina Ziekenhuis Eindhoven.
\end{credits}
%
%
%
\bibliographystyle{splncs04}
\bibliography{references}

\end{document}